\documentclass[authoryear]{elsarticle}

\usepackage[english]{babel}
\usepackage{fancyhdr} 
\usepackage{url} 
\usepackage{caption} 
\usepackage{subcaption} 
\usepackage{natbib} 
\usepackage{multirow} 
\usepackage{float} 
\usepackage{amssymb}
\usepackage[intlimits]{amsmath} 
\usepackage{amsthm} 
\usepackage[german]{nomencl}
\usepackage{booktabs} 
\usepackage{geometry}
\usepackage{enumitem}
\usepackage{bbm}

\makenomenclature
\usepackage[pdfborder={000},colorlinks=true,linkcolor=blue,citecolor=blue]{hyperref}
\usepackage{listings}
\usepackage[all]{xy}

\usepackage{color}

\definecolor{b}{rgb}{0,0,.8}	
\definecolor{g}{rgb}{0,.6,0}	
\definecolor{n}{rgb}{0,0,0}	
\definecolor{h}{rgb}{0.4,0.2,0.2}	
\definecolor{v}{rgb}{0.2,0.6,0}

\DeclareMathOperator*{\argmin}{arg\,min}

\newcommand{\rec}[1]{\textcolor{black}{#1}}

\begin{document}

\title{Sparsity and Stability  \\ for Minimum-Variance Portfolios}

\author{Sven~Husmann}
\ead{husmann@europa-uni.de}

\author{Antoniya~Shivarova}
\ead{shivarova@europa-uni.de}

\author{Rick~Steinert}
\ead{steinert@europa-uni.de}
\address{Europa-Universit\"at Viadrina, Gro\ss e Scharrnstra\ss e 59, 15230 Frankfurt (Oder), Germany}

\begin{frontmatter}
\begin{abstract}
The popularity of modern portfolio theory has decreased among practitioners because of its unfavorable out-of-sample performance. Estimation errors tend to affect the optimal weight calculation noticeably, especially when a large number of assets is considered. To overcome these issues, many methods have been proposed in recent years, although most only address a small set of practically relevant questions related to portfolio allocation. This study therefore sheds light on different covariance estimation techniques, combines them with sparse model approaches, and includes a turnover constraint that induces stability. We use two datasets -- comprising 319 and 100 companies of the S\&P 500, respectively -- to create a realistic and reproducible data foundation for our empirical study. To the best of our knowledge, this study is the first to show that it is possible to maintain the low-risk profile of efficient estimation methods while automatically selecting only a subset of assets and further inducing low portfolio turnover. Moreover, we provide evidence that using the LASSO as the sparsity-generating model is insufficient to lower turnover when the involved tuning parameter can change over time.
\end{abstract}
\end{frontmatter}

\section{Introduction and main idea} \label{Introduction}
\rec{The mean-variance portfolio optimization of \cite{Markowitz.1952} is still one of the most widely used approaches for selecting an optimal portfolio of assets with uncertain returns. To implement this approach in practice, one needs to estimate two sets of parameters -- expected asset returns and the covariances of asset returns -- which are traditionally estimated using the sample means and sample covariances of past returns, respectively. Unfortunately, the literature has extensively found that portfolios based on these estimates exhibit extremely poor out-of-sample performance, as the errors in the parameter estimates are carried over to the estimation errors in the portfolio weights.\footnote{See, for example, \citet{Michaud.1989}, \citet{Best.1991b}, \citet{Chopra.1993}, \citet{Broadie.1993}, \cite{Litterman.2003}, and \citet{DeMiguel.2009b, DeMiguel.2009}.} }

However, one portfolio on the efficient frontier that does not require researchers to estimate the mean decreases the estimation error: the minimum-variance portfolio. Interestingly, \citet{DeMiguel.2009} show that the mean-variance portfolio is outperformed out-of-sample not only by the naive portfolio but also by the minimum-variance portfolio in terms of the Sharpe ratio as well as the certainty-equivalent for most of their investigated datasets. This is a remarkable finding, as the minimum-variance portfolio specifically aims to reduce variance and not increase the mean.

In this study, we therefore introduce a sparse and stable model approach that focuses on the minimum-variance portfolio. We regard a portfolio as sparse if it selects a small number of assets out of a large investment space and as stable if its weights exhibit only small changes for each rebalancing step over time.

We show that recent advances in estimating covariance matrices have improved the risk profile of this portfolio type vastly. Nonetheless, we find that even the most accurate and recent covariance estimation techniques need to be updated to meet the common requirements of investors, especially in terms of transaction costs. For that, we focus on two specifications: a sparse selection of a subset of all assets and low turnover. This study is the first to show that combining highly efficient covariance estimators with penalty terms and turnover constraints can lead to portfolios that have the same low risk as their unconstrained counterparts, while simultaneously keeping a lower number of assets as well as lower turnover. We further provide new evidence on the usage of a mere penalized optimization for the weights toward achieving low turnover.

The remainder of this paper is organized as follows. After this section, we review the subject to provide insights from the scientific literature as well as from a practitioner's perspective. We then introduce three model setups. In the empirical study section, we describe our methodology, provide empirical evidence, and draw conclusions. A summary of our work and final remarks are provided in the last section.

\section{Review of the literature} \label{sec_review}
In recent decades, research on estimating covariance matrices under specific premises has gained increasing popularity. Researchers from different fields have adopted various strategies to tackle the many issues arising because of high dimensionality in the data and ill-conditioned covariance estimation. For instance, \citet{Bouchaud.2009}, \citet{Fan.2013}, and \citet{Ledoit.2017} develop methods based on random matrix theory to estimate covariance matrices. Shrinking approaches have also been established, which have in common that the covariance matrix estimated with maximum likelihood (ML) is mixed with one or many target matrices, as shown, for example, by \citet{Ledoit.2004}. Other researchers have focused more on the time dependency of returns; see, for instance, \citet{Engle.2017} for a recent study.

Owing to the increase in the transparency of research and advancement of computational power, implementing these models has become gradually easier for practitioners. Nonetheless, with the rise in data availability, investors seek diversification in large markets, but are limited by organizational as well as legal restrictions. The aforementioned estimation procedures, however, do not always make it straightforward to include these requirements. For example, the option to choose from a large set of stocks creates the problem of selection, as it is unfavorable for investors to hold a large number of stocks with a small relative weight \citep{lobo2007portfolio,takeda2013simultaneous}. This is mostly because of the fixed costs associated with including each asset. Further, the matter of transaction costs also plays a crucial role in portfolio choice. If, over time, the weights of the portfolio change too much and thus require frequent rebalancing, the investor faces unnecessarily high costs. To tackle these issues, researchers including \citet{konno2002portfolio} and \citet{lobo2007portfolio} have independently developed methods to include costs in the portfolio optimization itself. Another important, commonly applied restriction is the exclusion of short-sale positions, which is usually induced by law.

These \rec{real-world constraints are not only relevant because of legal and other regulatory circumstances. \citet{Jagannathan.2003}, for example, show that self-imposed short-sale constraints can significantly improve the out-of-sample performance of portfolios.} In particular, so-called norm constraints, or penalizing constraints, which help induce sparsity or shrinkage, can both reduce the amount of necessary assets as well as improve estimation accuracy. They are implemented by penalizing the $p$-vector norm of the asset weights with an additional factor, usually called $\delta$. \rec{Since \citet{Brodie.2009} and \citet{DeMiguel.2009b} introduced the $\ell_1$-norm and squared $\ell_2$-norm to portfolio optimization, these have gained increasing attention in this field of research. The $\ell_1$-norm is primarily applied to create sparse portfolios (i.e., portfolios with only a few active positions and, thus, lower overall estimation risk). The squared $\ell_2$-norm controls the balance of a portfolio, which can be measured by the deviation of its weights from the weights of an equally weighted portfolio.} Another norm tailor-made for subset selection is the $\ell_0$-norm, which overcomes the issue of the $\ell_1$-norm of deselecting potentially relevant assets \citep{takeda2013simultaneous}. Advances in this field have included, for instance, \citet{Fastrich.2014}, who compare standard cardinality constraints with different $\ell_p$-norms. Despite its high quality, however, the $\ell_0$-norm suffers from an absence of feasible solutions for estimating portfolios.

\section{Model setup}
To introduce our modeling approach, we start with the standard approach for constructing a global minimum-variance portfolio with $n$ assets and extend it further with restrictions to induce sparsity as well as stability. We assume that the investor uses a one-time optimization in the current period and readjusts his or her investment decision in subsequent periods by repeatedly executing the one-time optimization for the minimum-variance portfolio. The true covariance matrix $\Sigma$, which is unknown to the investor, needs to be estimated using the return data of $\tau$ historic time points.

\subsection{Standard minimum-variance portfolios} \label{sec_gmv}
To find the portfolio exhibiting the lowest variance among all assets, an investor faces the following optimization problem:
\begin{eqnarray}
\widehat{w} = \argmin_{w}
\label{eq_general_frame_full1}  
&& 
w' \widehat{\Sigma} w 
\label{eq_general_frame_objective} \\	
\text{s.t.}
&&
A w = a,  
\label{eq_general_frame_st_1a}
\end{eqnarray}
where $\widehat{w}$ is the estimated vector of portfolio weights, $\widehat{\Sigma}$ is the estimated covariance matrix, and \eqref{eq_general_frame_st_1a} represents the sum constraint. The latter means that all weights must sum to 1, as we choose $A=1_n$ and $a=1$, where $1_n$ is the $n$-dimensional vector of ones. We refer to this model as the Standard model.

\subsection{Sparse minimum-variance portfolios} \label{sec_sparse}
If a smaller number of assets is selected from the whole investment space, the optimization problem can be adjusted by adding an $\ell_1$-norm constraint, often referred to as the least absolute shrinkage and selection operator (LASSO), as follows:
\begin{eqnarray}
\widehat{w} = \argmin_{w}
\label{eq_general_frame_full2}  
&& 
w' \widehat{\Sigma} w 
 \\	
\text{s.t.}
&&
\, \,\,\,\,A w = a  
\label{eq_general_frame_st_1b} \\
&&
||w||_1^{\phantom{1}} \leq \delta. 
\label{eq_general_frame_st_2a}
\end{eqnarray}
Equation \eqref{eq_general_frame_st_2a} introduces a sparsity parameter $\delta$, which controls the shrinkage of the portfolio weights toward zero. Choosing a high $\delta$ value will lead to the same result as the standard optimization problem, whereas a sufficiently small $\delta$ will restrict the parameter space to a few assets. The solution to the optimization problem can still easily be found using standard quadratic programming with linear constraints, as it is possible to reallocate $w$ into its positive part $w^+=max(w,0)$ and $w^-=max(-w,0)$. The left-hand side of constraint \eqref{eq_general_frame_st_2a} can then be rewritten as $||w||_1^{\phantom{1}}={1_n}w^+ +{1_n}w^-$. The whole optimization problem then becomes
\begin{eqnarray}
\begin{bmatrix} \widehat{w}^+ \\ \widehat{w}^- \end{bmatrix}
= \argmin_{w^+, w^-}
&&
\begin{bmatrix} w^+ \\ w^- \end{bmatrix}^T
\begin{bmatrix} \widehat{\Sigma}, & -\widehat{\Sigma} \\
-\widehat{\Sigma} , & \widehat{\Sigma} \end{bmatrix}
\begin{bmatrix} w^+ \\ w^- \end{bmatrix} 
+ \begin{bmatrix} \lambda 1_n \\ \lambda 1_n \end{bmatrix}^T
\begin{bmatrix} w^+ \\ w^- \end{bmatrix}
\nonumber \\	
\text{s.t.}
&&
\begin{bmatrix} A, -A \end{bmatrix} 
\begin{bmatrix} w^+ \\ w^- \end{bmatrix} = a \text{ and }  
\begin{bmatrix} 0_n \\ 0_n \end{bmatrix}
\leq \begin{bmatrix} w^+ \\ w^- \end{bmatrix},
\label{eq_prop_L1_L2_QP}
\end{eqnarray}
which is a quadratic optimization with linear constraints and a Lagrange parameter $\lambda$. Owing to the combination of \eqref{eq_general_frame_st_1b} and \eqref{eq_general_frame_st_2a} the parameter space cannot be fully restricted (e.g., to up to zero assets).

\subsection{Sparse and stable minimum-variance portfolios} \label{sec_sparse_stable}
In a real-world application of a minimum-variance portfolio, investors are prone to costs, which are related to the turnover of the portfolio (e.g., transaction costs). The introduction of a shrinkage-type constraint such as \eqref{eq_general_frame_st_2a} can to some degree account for transaction costs, as the parameter $\delta$ will penalize high asset weights $w_i$ and therefore indirectly reduce the possibility of vast changes between the subsequent rebalancing time points. However, we argue that the LASSO alone cannot sufficiently decrease turnover, as it has no information on past weights. Although the weights overall are relatively small, if many of these change at a time or many weights that were earlier deselected (i.e., set to zero) are now selected, turnover might still be reasonably high. Hence, we include a specific turnover constraint that works as a proxy for transaction costs. The optimization problem now changes to

\begin{eqnarray}
\widehat{w} = \argmin_{w}
\label{eq_general_frame_full3}  
&& 
w^T \widehat{\Sigma} w 
 \\	
\text{s.t.}
&&
\, \,\,\,\,A w = 1  
 \\
&&
||w||_1^{\phantom{1}} \leq \delta \\
&&
\,\,\,\,\,\,\,\,\,\,w \leq k + w_a
\label{eq_general_frame_turn1}  \\
&&
\,\,\,\,\,\,-w \leq k - w_a, 
\label{eq_general_frame_turn2}
\end{eqnarray}

where the stability-inducing constraints \eqref{eq_general_frame_turn1} and \eqref{eq_general_frame_turn2} respectively form the turnover constraints dependent on the weights of the previous optimization step $w_a$ and a tuning parameter $k$, which controls the allowed change in the positions for one rebalancing step in both directions. The usual turnover constraint $|w|\leq w_a+k$ is rewritten as \eqref{eq_general_frame_turn1} and \eqref{eq_general_frame_turn2} to avoid using non-linear constraints. With respect to the adjustment of $w$ to $w^+$ and $w^-$, as explained before, the whole optimization problem remains a quadratic problem with linear constraints and is therefore efficiently solvable with standard optimization software. We refer to this model as the LASSO with the turnover constraint model.\footnote{Setting $k=\infty$, models \eqref{eq_general_frame_full3} and \eqref{eq_general_frame_full2} become the same and when $k=\infty$ and $\delta=\infty$, models \eqref{eq_general_frame_full3} and \eqref{eq_general_frame_full1} yield the same results.}

\subsection{Model discussion}
In theory, estimating the global minimum-variance portfolio, as introduced in Section \ref{sec_gmv}, should be sufficient to obtain the portfolio with the lowest variance and, thus, the lowest risk. The restriction introduced in Section \ref{sec_sparse} would then become obsolete, as no matter how large the investment space is, including assets will always reduce or keep the variance at the same level, but never increase it. The additional constraints \eqref{eq_general_frame_turn1} and \eqref{eq_general_frame_turn2} would then function as merely a practitioners' constraint by reducing turnover and thus, transaction costs.

However, as pointed out earlier, the covariance matrix needs to be accurately estimated to obtain optimal results. As researchers have pointed out, even small estimation differences can lead to vast deviations from the true efficient frontier; see, for instance, \citet{Jobson.1980, Jobson.1981b}, and \citet{Frost.1986, Frost.1988} for some of the earliest studies of this topic. Moreover, \citet{Kan.2007} argue that the unbiased ML estimator for $\Sigma$ has unwanted properties for specific ratios $\frac{n}{\tau}$, even when the underlying return data follow a normal distribution. 

Several methods have been proposed to reduce the estimation error for the covariance matrix, most applying some form of shrinkage (e.g., \citet{Ledoit.2004}). Nonetheless, as any estimation error in the covariance matrix directly influences the estimation of the weights $w$ of the portfolio, some authors shrink the weights $w$ directly by, for instance, combining it with another portfolio. For instance, \citet{Tu.2011} apply a mean-variance portfolio combined with an equally weighted portfolio. \citet{Jagannathan.2003} even argue that any constraint on the optimization procedure might help reduce the estimation error.

All the model approaches introduced in Sections \ref{sec_gmv}--\ref{sec_sparse_stable} can therefore reduce the estimation error in their own way. To choose a suitable standard minimum-variance portfolio, we use highly efficient and recent estimators for the covariance matrix $\Sigma$. The constraint \eqref{eq_general_frame_st_2a} of our sparse model will not only create sparsity but also directly reduce the estimation error in the weights $w$, as these will be shrunk toward zero. In our sparse and stable model presented in Section \ref{sec_sparse_stable}, we further stabilize the portfolio estimations by introducing the turnover constraint. This directly forces the weights to change only in a small window of length $2k$, and hence indirectly applies another way of prohibiting estimation errors due to potential misspecifications in the data.

Overall, this framework allows us to study the behavior of the LASSO, one of the most common sparsity-inducing methods, when the covariance estimate has already adjusted to potential problems. We further check whether the common assumption that the LASSO can reduce turnover significantly on its own continues to hold true when the covariance matrix is already estimated sufficiently (e.g., \citet{Brodie.2009}) By doing so, we can gain insights into how the introduction of a common turnover constraint changes the risk profile of these portfolios.

\section{Empirical study}
As our results are solely based on an empirical analysis, it is crucial to employ a suitable empirical setup to ensure the validity and reproducibility of our findings. As an investor that uses a minimum-variance portfolio seeks, by definition, the portfolio exhibiting the lowest variance, our empirical study must reflect this objective. Indeed, all the investigated portfolios include so-called tuning parameters -- parameters important for the optimization procedure but not yet optimized by theoretical analysis. Instead, one of these parameters is identified by cross-validation in combination with machine power, whereas the other will be set to a constant value.

In our case, we have two tuning parameters: $\delta$, resulting from the LASSO constraint, and $k$, emerging from the turnover constraint. 
Owing to computational restrictions, we use $\delta$ as the tuning parameter to achieve the lowest variance and therefore optimize its value with cross-validation. The parameter $k$ of the turnover constraint is kept constant throughout the dataset, as we only use it to reduce turnover compared with the unconstrained benchmark.

Hence, we do not set the value of $\delta$ so that it meets specific well-known constraints such as the short-sale constraint (see \citet{DeMiguel.2009b}). In contrast to other authors such as \citet{zhao2019risk}, we do not want to achieve any practitioner's rule for investment and therefore do not keep $\delta$ as a constant, independent of the present market situation. By contrast, we allow the $\delta$ value to change in every period, as we always want to achieve the optimization goal, that is, finding the portfolio with the lowest variance. This, in our opinion, a more realistic approach leads to a more likely change in the chosen assets and higher turnover. This in turn provides another reason for imposing an additional turnover constraint, as in model \eqref{eq_general_frame_full3}.

\subsection{Data}
For our empirical study, we use S\&P 500 stock price data from the Thomson Reuters EIKON database. This covers daily data from January 1998 to the end of December 2018, with $T=5282$ observations overall. Our analysis is based on discrete returns, calculated as $r_t=\frac{P_t-P_{t-1}}{P_{t-1}}$. To avoid transforming the data and therefore potentially distorting valuable information, we only focus on those stocks present throughout the data period (319 stocks). To check whether dimensionality influences our results, we analyze the surviving 319 companies as well as a randomly generated subset of 100 stocks of the original 319. All our models are estimated by taking into account the returns of approximately the past two years of trading (i.e., $\tau=2*252=504$ observations). This results in 4778 trading days of out-of-sample returns from our different model approaches.

\begin{figure}[h!]
\centering
\begin{subfigure}{.5\textwidth}
  \centering
  \includegraphics[width=.99\linewidth]{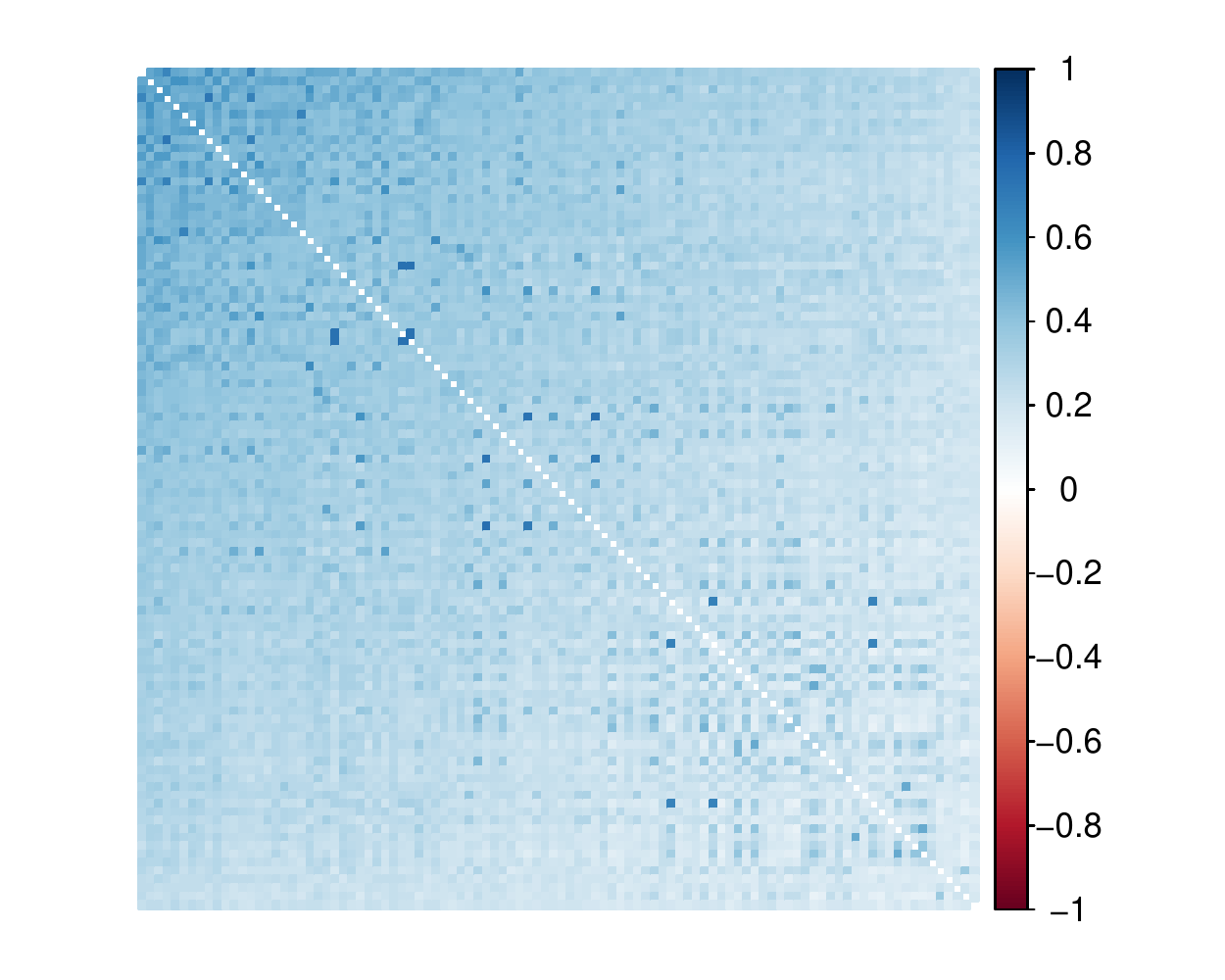}
  \caption{100 stocks}
  \label{fig:sub1}
\end{subfigure}%
\begin{subfigure}{.5\textwidth}
  \centering
  \includegraphics[width=.99\linewidth]{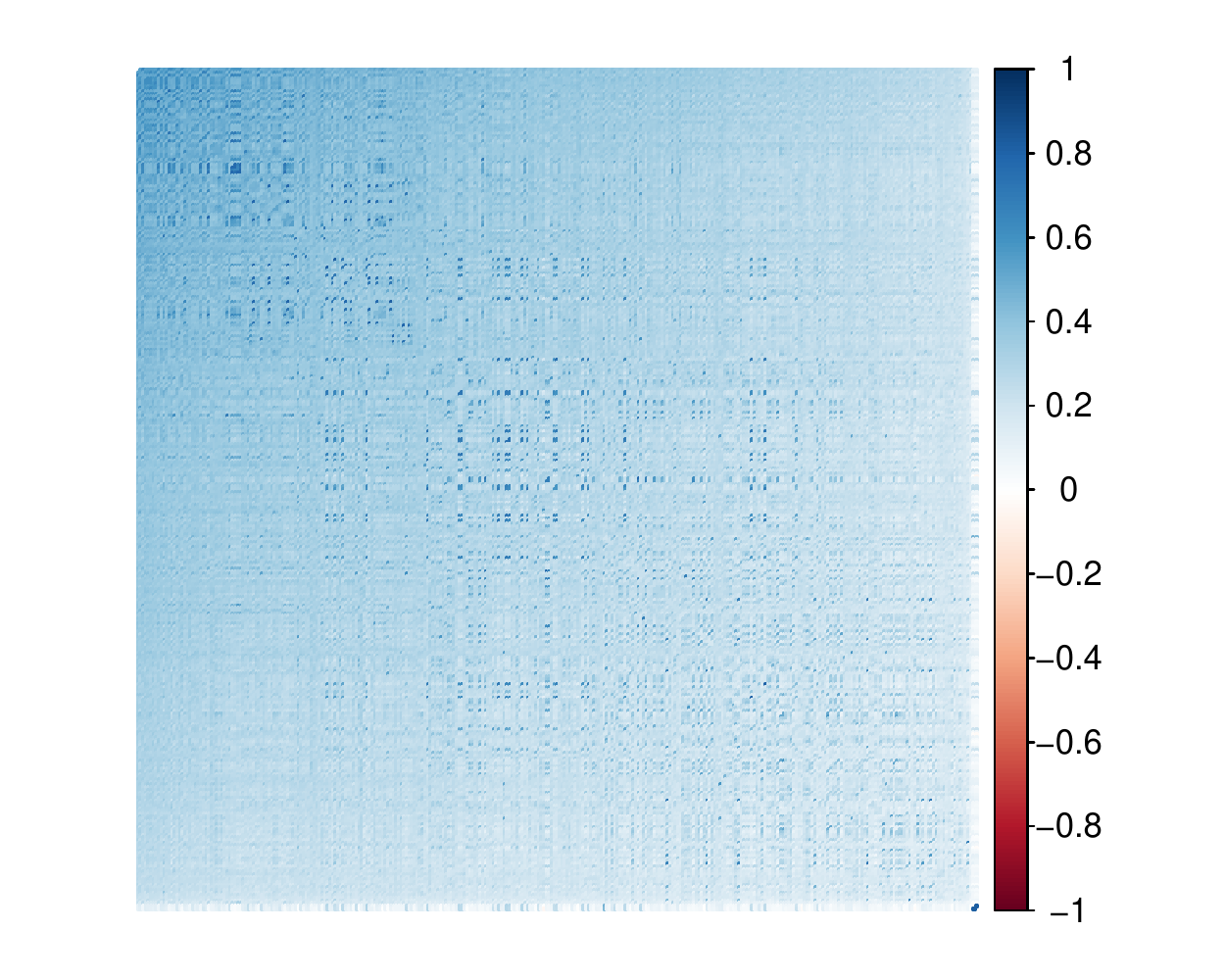}
  \caption{319 stocks}
  \label{fig:sub2}
\end{subfigure}
\caption{Correlation plots for the stocks of our two datasets. The blue correlation indicates a strong positive correlation; white, slight to no correlation; and red, a strong negative correlation.}
\label{fig:correlationplot}
\end{figure}

To illustrate the structure of our two datasets, Figure \ref{fig:correlationplot} presents their correlation plots. This figure shows the sample correlation of each possible pair of stock returns. As all the correlations are close to zero or positive for both datasets, we order the stocks according to their first principal component.\footnote{In this way, we can easily display the magnitude of the correlation structure without checking, for instance, 50,721 correlations unit by unit, as would be needed in the case of 319 stocks.} \citet{friendly2002corrgrams} and \citet{wei2017package} provide examples of an R-implementation. Both figures show that a large number of stocks exhibit strong correlations with each other. However, from the top left to bottom right of the figures, the overall correlation diminishes to a slight positive correlation, in some cases even close to zero. Further, the random selection of 100 out of the 319 stocks does not visually break the underlying correlation structure of the data, as both plots seem to have a similar appearance.

\subsection{Variance estimators}
To thoroughly analyze whether the introduced LASSO and turnover constraints decrease the number of assets as well as turnover, while maintaining a low-variance profile, we use some recent and efficient variance estimators. Starting with one of the most commonly used estimators among practitioners and researchers, we calculate the sample covariance estimator, defined as
\begin{eqnarray*} 
\widehat{\Sigma}_\mathit{S}=\frac{1}{\tau-1} \left(R - \widehat{\mu} 1' \right) \left(R -  \widehat{\mu} 1' \right)',
\end{eqnarray*}
where $R \in \mathbb{R}^{n \times t}$ is the matrix of past returns and $\widehat{\mu} \in \mathbb{R}^{n}$ the vector of expected returns (here, estimated as average returns).  
At high concentration ratios, $q=n/\tau\rightarrow 1$, the empirical variance, although unbiased, exhibits high estimation variance and, therefore, a high out-of-sample estimation error. To minimize this estimation error, a linear shrinkage procedure can be applied to the unbiased sample estimator by combining it with a target covariance matrix. Following \citet{Ledoit.2003}, the variance estimator becomes
\begin{eqnarray}\label{eq_covest_lwlin}
\widehat{\Sigma}_\mathit{LW_L}  = s\widehat{\Sigma}_\mathit{T} + (1-s)\widehat{\Sigma}_\mathit{S},
\end{eqnarray}
where $\widehat{\Sigma}_T$ is the estimate of a specific target covariance matrix and $s$ is a shrinkage constant with $s \in [0,1]$. Assuming identical pairwise correlations between all $n$ assets, the target matrix is substituted with the constant covariance matrix as in \citet{Ledoit.2004}.

A more sophisticated shrinking method is non-linear shrinkage, as suggested by \citet{Ledoit.2017}. As this estimator shrinks the eigenvalues individually; small, potentially underestimated eigenvalues are pushed up, while large, potentially overestimated eigenvalues are pulled down. Without going into further detail, we write the non-linear shrinkage estimator as
\begin{eqnarray}\label{eq_covest_lwnonlin}
\widehat{\Sigma}_\mathit{LW_\mathit{NL}} = V\widehat{E}_\mathit{LW_\mathit{NL}} V',
\end{eqnarray}
where $V$ is the matrix of the orthogonal eigenvectors and $\widehat{E}_\mathit{LW_\mathit{NL}}$ is the diagonal matrix of the shrunk eigenvalues, as shown by \citet{Ledoit.2012, Ledoit.2015}. Because $\widehat{\Sigma}_\mathit{LW_\mathit{NL}}$ is proven to be asymptotically optimal within the class of rotationally equivariant estimators, we might expect it to perform better than any of the aforementioned estimators, especially in cases of large concentration ratios.

Hence, we further extend our analysis to factor-based covariance estimation methods, which assume a specific structure in the covariances of asset returns. One promising example of that family of variance estimators is the principal orthogonal complement thresholding (POET) estimator provided by \citet{Fan.2013}. Here, the principal components of the sample covariance matrix $\widehat{\Sigma}_\mathit{S}$ are used as factors. Moreover, subsequent adaptive thresholding with a threshold parameter $\theta$ is applied to the covariance of the residuals of the estimated factor model \citep[see, e.g.,][]{Cai.2011}.\footnote{For the application in our study, we use the R-code provided by \citet{Fan.2013} within the R-package $POET$ and repeatedly apply a separate cross-validation to obtain the number of factors $K$, as suggested by the authors.} Therefore, the POET estimator has the form:
\begin{eqnarray}\label{eq_covest_poet}
\widehat{\Sigma}_\mathit{POET}=\sum^K_{i=1}\widehat{\xi}_i v_i v_i' + \widehat{\Sigma}^{\theta}_{u,K},
\end{eqnarray}
where $v_i$ is the eigenvector to asset return $i$, $\xi_i$ is the corresponding eigenvalue, and $\widehat{\Sigma}^{\theta}_{u,K}$ is the idiosyncratic covariance matrix after the applied thresholding procedure with threshold level $\theta$.

In particular, the estimators \eqref{eq_covest_lwnonlin} and \eqref{eq_covest_poet} estimate the global minimum-variance portfolios well, and thus, they can be considered to be the state-of-the-art among homoscedastic variance estimators for return data.

\subsection{Performance measures}
\rec{To evaluate the out-of-sample performance of each portfolio, we report the out-of-sample variance, average daily turnover,} average number of included assets, and average number of short sales as
\begin{eqnarray} 
\text{Standard Deviation} &=& \frac{1}{T-\tau}\sum^{T-1}_{t=\tau}\left(w_t' r_{t+1} - \widehat{\mu}\right)^2
\nonumber \\
\text{Turnover} &=& \frac{1}{T-\tau-1}\sum^{T-1}_{t=\tau+1}\sum^{n}_{j=1}\left(\left|w_{j, t+1}-w_{j,t^+}\right|\right)
\label{eq_perf_measures_turnover}
\\
\text{Average Assets} &=& \frac{1}{T-\tau}\sum^{T}_{t=\tau+1}\sum^{n}_{j=1}\mathbbm{1}_{\{w_{j,t}\neq 0\}}
\label{eq_perf_measures_avg_assets}
\\
\text{Average Short sales} &=& \frac{1}{T-\tau}\sum^{T}_{t=\tau+1}\sum^{n}_{j=1}\mathbbm{1}_{\{w_{j,t}<0\}},
\label{eq_perf_measures_avg_short sales}
\end{eqnarray}
\rec{where $w_t$ are the portfolio weights chosen at time $t$, $w_t' r_{t+1}$ is the out-of-sample portfolio return, $\widehat{\mu}= \frac{1}{T-\tau}\sum^{T-1}_{t=\tau}w_t' r_{t+1}$ is the out-of-sample expected return, $w_{j,t^+}$ denotes the portfolio weight in asset $j$ before rebalancing at $t+1$} but scaled back to sum to 1 \rec{and $w_{j, t+1}$ is the portfolio weight in asset $j$ after rebalancing at $t+1$.}

Since we consider a variance minimization problem, daily out-of-sample portfolio variance is of utmost importance. Hence, we check whether the calculated out-of-sample variance of the LASSO-based method in \eqref{eq_general_frame_full2} as well as the LASSO and turnover-based method in \eqref{eq_general_frame_full3} have significantly different standard deviations than their standard counterpart in \eqref{eq_general_frame_full1}. Therefore, \rec{we perform the two-sided HAC test with the Parzen kernel for the differences in variances, as described by \citet{Ledoit.2008}, and report the corresponding $p$-values. In accordance with the literature on portfolio optimization and estimation risk reduction, we use average daily turnover as a proxy for the arising transaction costs \citep[e.g.,][]{DeMiguel.2009b, Dai.2018}.} 

We next evaluate the portfolio composition with respect to the number of non-zero investments and short sales as well as the development of the short-sale budget over time. Finally, we analyze the final $\delta$ values of model types \eqref{eq_general_frame_full2} and \eqref{eq_general_frame_full3}. All the values are reported on a daily basis.


\subsection{Course of action}
For our empirical work, we use a non-expanding rolling window study that incorporates cross-validation for our tuning parameter. As mentioned earlier, we evaluate the tuning parameter $\delta$ for the LASSO constraint, so that it may change each day. The parameter $k$ is left constant over time, set to $0.0005$ for the 319 S\&P dataset and $0.001$ for the 100 S\&P dataset. These values were found by checking different values $k$ for each dataset in a small subsample. Changing $k$ by a reasonably large number did not result in vastly different outcomes. In general, choosing a too large value for $k$ leads to a portfolio that still has high turnover, whereas choosing it to be too small worsens its risk/return profile. The more the assets considered, the lower $k$ should be.

To analyze the impact of the LASSO constraint model and LASSO with the turnover constraint model, we implement a simple one-fold cross-validation for the tuning parameter $\delta$. However, because of the described model representation of \eqref{eq_prop_L1_L2_QP}, which allows us to simplify the absolute value constraint, we apply our cross-validation toward the $\lambda$ Lagrange parameter instead of $\delta$ and restore all $\delta$ values in a second step by simply calculating $||w||_1^{\phantom{1}}=\delta$. 

We start with $t=1$, January 2, 1998, and use the following daily returns up to $t=504$ to create an in-sample dataset covering approximately two years of daily returns. From that data sample, we take another smaller subsample for our cross-validation consisting of the first $504-20=484$ observations. We then calculate models \eqref{eq_general_frame_full2} and \eqref{eq_general_frame_full3} using 20 different $\lambda$ values chosen from a linear sequence of numbers from $\lambda_{t+1}=\lambda_{t}+0.00001$ to $0$, whereas we initialize $\lambda_1$ with $0.00001$. The resulting weights of these 20 models are then applied to the first subsequent daily return of the cross-validation subsample (here, the $485th$ observation) to create an individual daily portfolio return for both models. The subsample is then shifted by one and the procedure carried out again. This is repeated until we reach the 20 observations we previously omitted. We then compare the standard deviations of the 20 out-of-sample cross-validation returns for each $\lambda$ for both models individually. After receiving an optimal $\lambda$, chosen to be that corresponding to the lowest standard deviation, we set our final $\lambda$ to be $\lambda_t$ for each model individually. Next, we calculate the weights using models \eqref{eq_general_frame_full2} and \eqref{eq_general_frame_full3} for all the selected data on daily in-sample returns. The true out-of-sample returns are then constructed by multiplying the calculated weights by the returns of the following period (here, the $505th$ observation). As model \eqref{eq_general_frame_full1} needs no cross-validation, we calculate it only at this point to receive its out-of-sample portfolio return as well. We then proceed by shifting the former in-sample data by one period (i.e., a day) and repeat the procedure 4778 times until the last out-of-sample daily return covers December 31, 2018.

\section{Results}
After applying the abovementioned techniques to the 319 and 100 S\&P 500 stock returns, we can analyze and compare the portfolio strategies. Table \ref{tab_319} shows the main results for our first dataset with 319 stocks.

\begin{table}[h!]
\begin{tabular}{  c  c  c  c  c  } 

	 & ML & \citet{Ledoit.2003} & \citet{Ledoit.2017} & \citet{Fan.2013} \\ \hline
	\multicolumn{5}{c}{Standard deviation p.a.}  \\ \hline
	Standard & 0.1387 & 0.1064 & 0.1004 & \underline{\textbf{0.0974}} \\ 
	LASSO & 0.1025 & \textbf{0.0990} & 0.1000 & 0.0975 \\ 
	 & (0.0000) & (0.0000) & (0.5392) & (0.8979) \\ 
	LASSO $+$ TO & \textbf{0.0998} & 0.0991 & \textbf{0.0997} & 0.0981 \\
	 & (0.0000) & (0.0000) & (0.5106) & (0.4153) \\ \hline
	
\multicolumn{5}{c}{Turnover}  \\ \hline
	Standard & 0.7856 & 0.3279 & 0.1757 & 0.1472 \\ 
	LASSO & 0.4125 & 0.3979 & 0.3062 & 0.2811 \\
	LASSO $+$ TO & \textbf{0.1439} & \textbf{0.1379} & \textbf{0.1392} & \underline{\textbf{0.1215}} \\ \hline
	\multicolumn{5}{c}{Average assets}  \\ \hline
	Standard & 319.00 & 319.00 & 319.00 & 319.00 \\ 
	LASSO & \underline{\textbf{120.19}} & \textbf{144.49} & \textbf{188.78} & \textbf{193.97} \\ 
	\textit{Percentage of full} & \textit{37.68} & \textit{45.29} & \textit{59.18} & \textit{60.80} \\ 
	LASSO $+$ TO & 265.22 & 255.53 & 267.06 & 233.57 \\ 
	\textit{Percentage of full} & \textit{83.14} & \textit{80.10} & \textit{83.72} & \textit{73.22} \\ \hline
		\multicolumn{5}{c}{Average short sales}  \\ \hline
	Standard & 152.26 & 154.30 & 133.11 & 141.73 \\ 
	LASSO & \underline{\textbf{44.94}} & \textbf{60.87} & \textbf{67.29} & \textbf{77.81} \\ 
	\textit{Percentage of full}  & \textit{14.09} & \textit{19.08} & \textit{21.09} & \textit{24.39} \\ 
	LASSO $+$ TO & 125.97 & 122.37 & 117.45 & 103.73 \\ 
	\textit{Percentage of full} & \textit{39.49} & \textit{38.36} & \textit{36.82} & \textit{32.52} \\ \hline
	 \end{tabular} 
	 \caption{Standard deviation p.a., average turnover per day, average assets as the mean of all non-zero weights, and average short sales as the mean of all weights greater 0 for the different models applied to the 319 S\&P dataset. The column headers represent the used variance estimation technique and rows report the results of the GMV calculated with \eqref{eq_general_frame_full1} (Standard), \eqref{eq_general_frame_full2} (LASSO), or \eqref{eq_general_frame_full3} (LASSO$+$TO). The results in bold represent the best model of that column for a specific measure and the underlined number represents the best overall model for that measure. The numbers in brackets represent the $p$-values of the test $H0: \sigma_{LASSO}=\sigma_{Standard}$ and $H0: \sigma_{LASSO+TO}=\sigma_{Standard}$, respectively. All the results are calculated based on the whole out-of-sample period consisting 4778 daily observations.}
	 \label{tab_319}
	 \end{table}

This table provides information on the four major aspects of our study, using the variance estimation methods in the columns and three modeling approaches in the rows. All the values are based on out-of-sample data. The first aspect is standard deviation p.a., which is the most important measure for minimum-variance portfolios. The second is average turnover per day as a proxy for transaction costs. Third, we report average assets, namely the average number of assets with weights different to zero, which were included in the daily final portfolio choice. Finally, the average amount of short sales per day is calculated as the average of all negative weights. For a better overview of the data, every number in bold corresponds to the model that performed best under a certain variance estimation technique and performance measure. The underlined numbers represent the overall best model in that performance category.

In terms of standard deviation, the Standard approach leads to the highest variances of all the estimation techniques, with only one exception (i.e., the POET estimator). The difference between the Standard approach and the LASSO approach with POET is, however, not significant, with a p-value of almost 90 percent. Nonetheless, for estimators such as ML and \citet{Ledoit.2003}, LASSO-based models can significantly decrease the variance, as indicated by a $p$-value of almost 0. This finding is in accordance with that of \citet{Dai.2018}, who conduct a similar study combining the estimator of \citet{Ledoit.2003} with LASSO and compare it with the short sale-constrained global minimum-variance portfolio. Surprisingly, the LASSO-based model with the linear shrinkage of Ledoit and Wolf performs even better than their non-linear version (\citet{Ledoit.2017}), independent of the model with which it is combined. Moreover, the POET estimator strictly dominates all the other variance estimators irrespective of whether a LASSO-based model is used. One possible explanation is that POET is the only covariance estimation method to incorporates an underlying factor model, a popular theory in finance for explaining the cross-section of returns. In general, the introduction of the LASSO with the turnover constraint model does not noticeably weaken the results for variance compared with the LASSO model. Hence, including the LASSO and LASSO with the turnover constraint models retains the low-variance profile of the Standard approach and can even sometimes significantly outperform it, providing suggestive evidence that incorporating the LASSO might improve variance overall as well.

\begin{figure}[h!]
\centering
\begin{subfigure}[b]{0.9\textwidth}
   \includegraphics[width=1\linewidth, height=8cm]{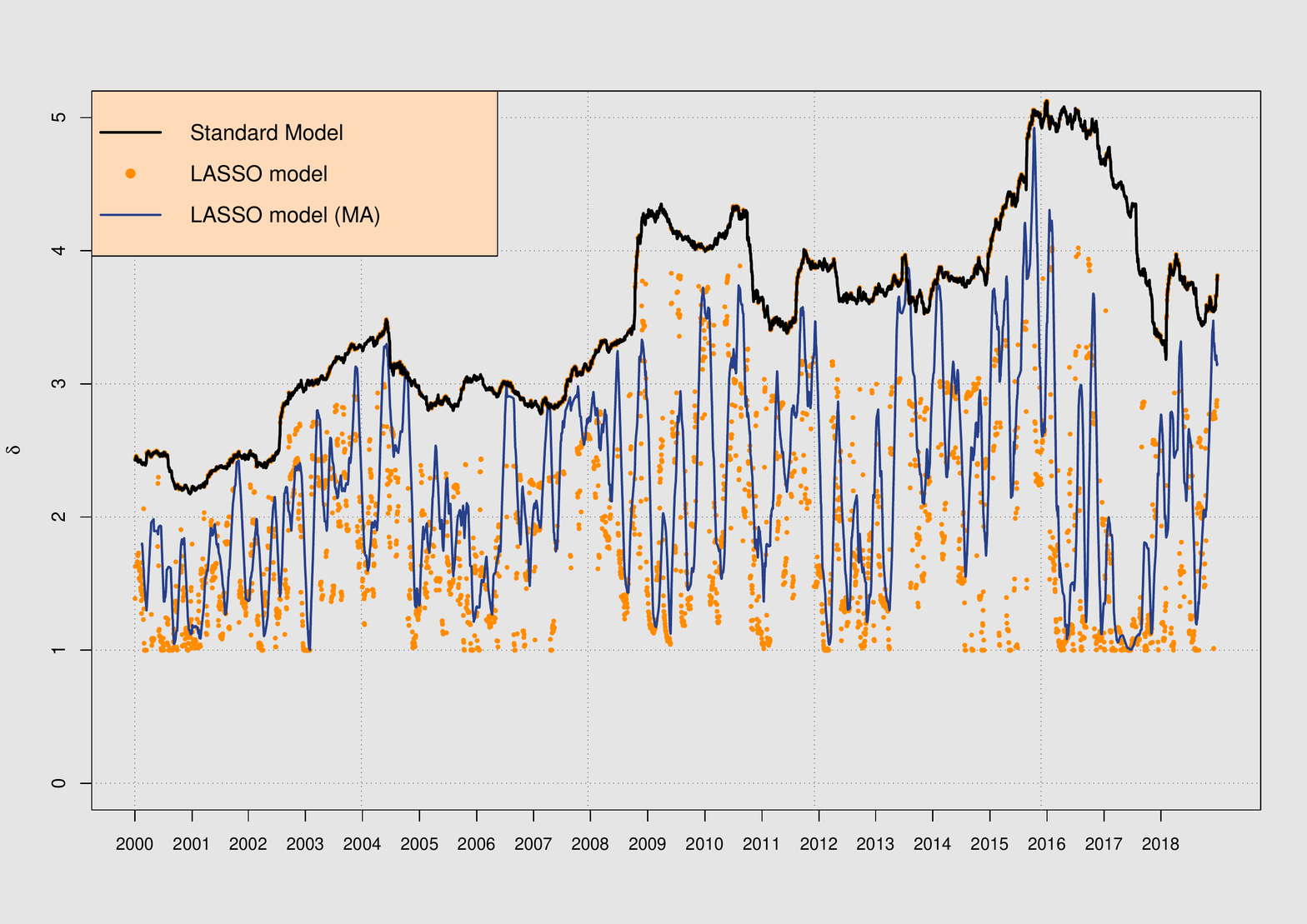}
  \caption{Calculated $\delta$ values for the LASSO and Standard model.}
  \label{fig:sub1_delta}
\end{subfigure}%

\begin{subfigure}[b]{0.9\textwidth}
  \includegraphics[width=1\linewidth, height=8cm]{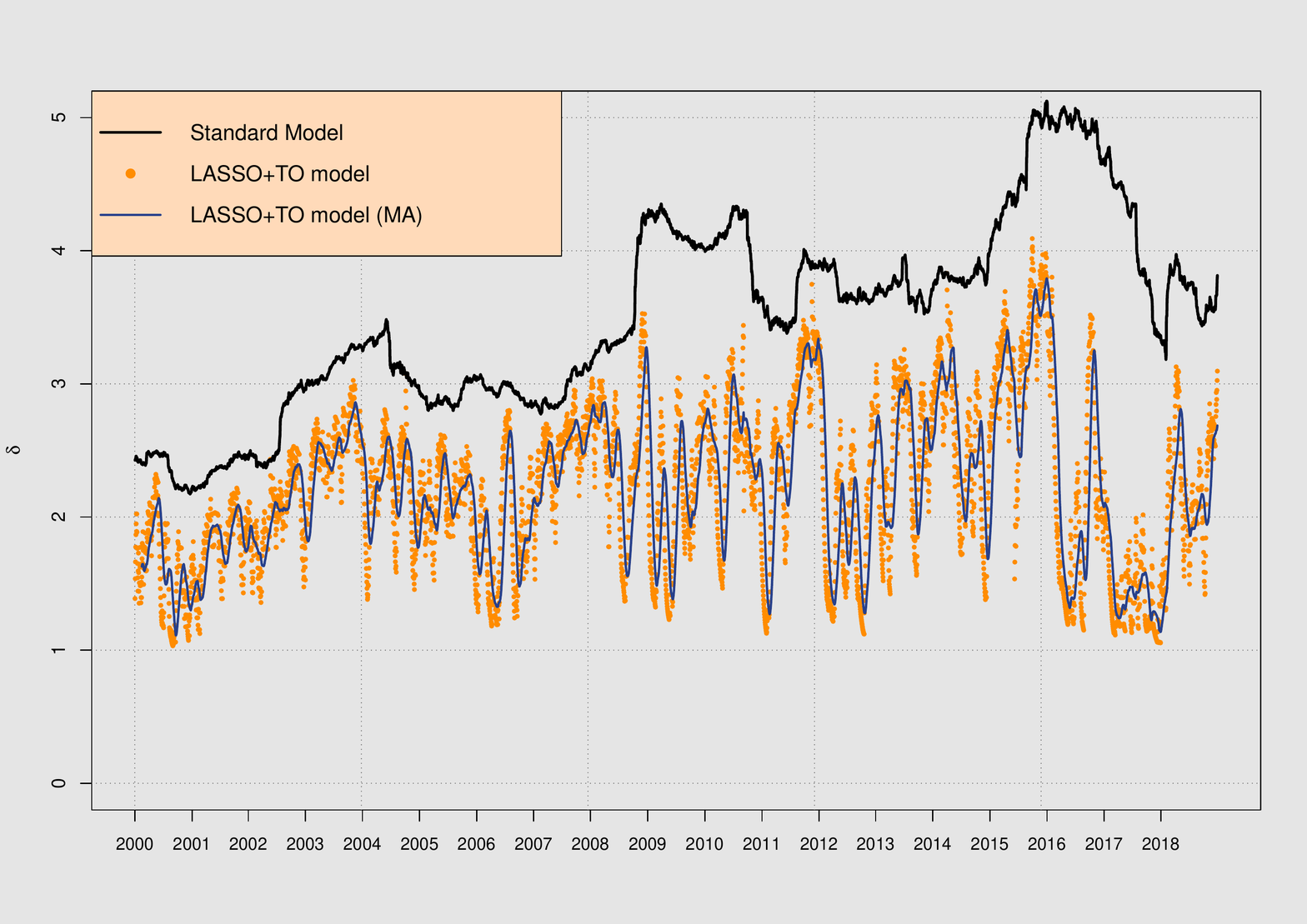}
  \caption{Calculated $\delta$ values for the LASSO with the turnover constraint and Standard model.}
  \label{fig:sub2_delta}
\end{subfigure}
\caption{Out-of-sample calculated $\delta$ values for the GMV portfolio types over time, where variance is estimated following \citet{Ledoit.2017}.}
\label{fig:deltaplot_nonlin}
\end{figure}

In terms of turnover, calculated in \eqref{eq_perf_measures_turnover}, the findings are surprising and novel. The table suggests that LASSO models without the turnover constraint have higher turnover than the Standard model, except for the ML estimator. Further, the more efficient a variance estimation method, the larger is the gap between the LASSO and Standard models in terms of turnover. Moreover, the gap between the LASSO and Standard model in terms of turnover is larger when the variance estimation is more efficient. Only for the poorly performing ML estimator do our findings coincide with those of, for example, \citet{Brodie.2009}. This finding confirms that LASSO models reduce turnover. Only when introducing the respective constraint, as in model \eqref{eq_general_frame_full3}, does turnover vastly decrease to levels below that of their Standard counterparts. This holds true for all the variance estimators. To support our findings and provide possible explanations, Figure \ref{fig:deltaplot_nonlin} provides an overview of the $\delta$ values of the portfolios over time, demonstrating the variance estimation technique of \citet{Ledoit.2017}.

Subfigure \ref{fig:sub1_delta} shows the calculated $\delta$ values for the Standard model as well as the LASSO model over time. The parameter $\delta$ in equation \eqref{eq_general_frame_st_2a} corresponds to the sum of absolute weights. While this parameter can only be calculated afterward in the Standard model, it is directly linked to the most influential tuning parameter in the LASSO model: the $\lambda$ Lagrange parameter. $\delta$ also provides information on the short-sale budget as well as practitioners' investment rules, as shown by \citet{zhao2019risk}. Naturally, as represented by the black line of Standard models' $\delta$, this parameter is stable over a few months but can have high fluctuations over years. In that sense, $\delta$ varies between 2.5 and 5 throughout the out-of-sample period. The LASSO model, however, does not incorporate these high levels of $\delta$, and thus in general has a lower short-sale budget than its Standard counterpart. This can be seen by the orange dots, but even better by the blue line, which represents the simple moving average of 30 days for these orange dots. The moving average is only applied for visualization purposes to show the variation in $\delta$.

In general, two main findings come from Subfigure \ref{fig:sub1_delta}. First, the overall short-sale budget with LASSO is always lower than or equal to that of the Standard portfolio approach, which coincides with the findings in the literature (e.g. \citet{Brodie.2009}). Second, the optimal $\delta$ for LASSO portfolios seems to be much more volatile than that for the Standard portfolio. In particular, the second finding causes the turnover of these portfolios to be high. If the chosen optimal $\lambda$ of each cross-validation out-of-sample step needs to be adjusted for every new step to minimize the standard deviation, the portfolio weights fluctuate more than usual. In studies focusing on LASSO-restricted portfolios, this is usually overseen, as most research assumes a stable $\lambda$ and, thus, a stable $\delta$ over time (e.g. \citet{zhao2019risk}).

Subfigure \ref{fig:sub2_delta} provides more evidence for this argument. The image illustrates the same characteristics, but now for the LASSO with the turnover constraint model. The orange dots seem to be much closer to each other than before. In addition, the blue line is far less volatile. This higher stability of $\delta$ over time leads to more stable weights and, thus, lower turnover.

Table \ref{tab_319} provides more information on the average number of assets over time, calculated as shown in \eqref{eq_perf_measures_avg_assets}. Naturally, the Standard GMV portfolio includes all 319 assets, as no restriction is imposed. However, the LASSO and LASSO with the turnover constraint portfolios both reduce the number of stocks in the portfolio. Again, for the unconstrained LASSO model, the number of included assets increases when a more efficient estimation technique is used. For the worst technique in terms of standard deviation (i.e., ML), the LASSO only selects 37.68 percent of all stocks on average, whereas this is 60.80 percent for the POET estimator of \citet{Fan.2013}. Imposing a turnover constraint on the LASSO changes these results slightly, meaning that the number of included assets increases compared with the regular LASSO. However, compared with the Standard model, the LASSO with the turnover constraint model is still strictly dominating in terms of sparsity, as it reduces the number of included assets to less than 84 percent of the whole asset universe for all cases.

\begin{figure}[h!]
\centering
\begin{subfigure}[b]{0.9\textwidth}
   \includegraphics[width=1\linewidth, height=8cm]{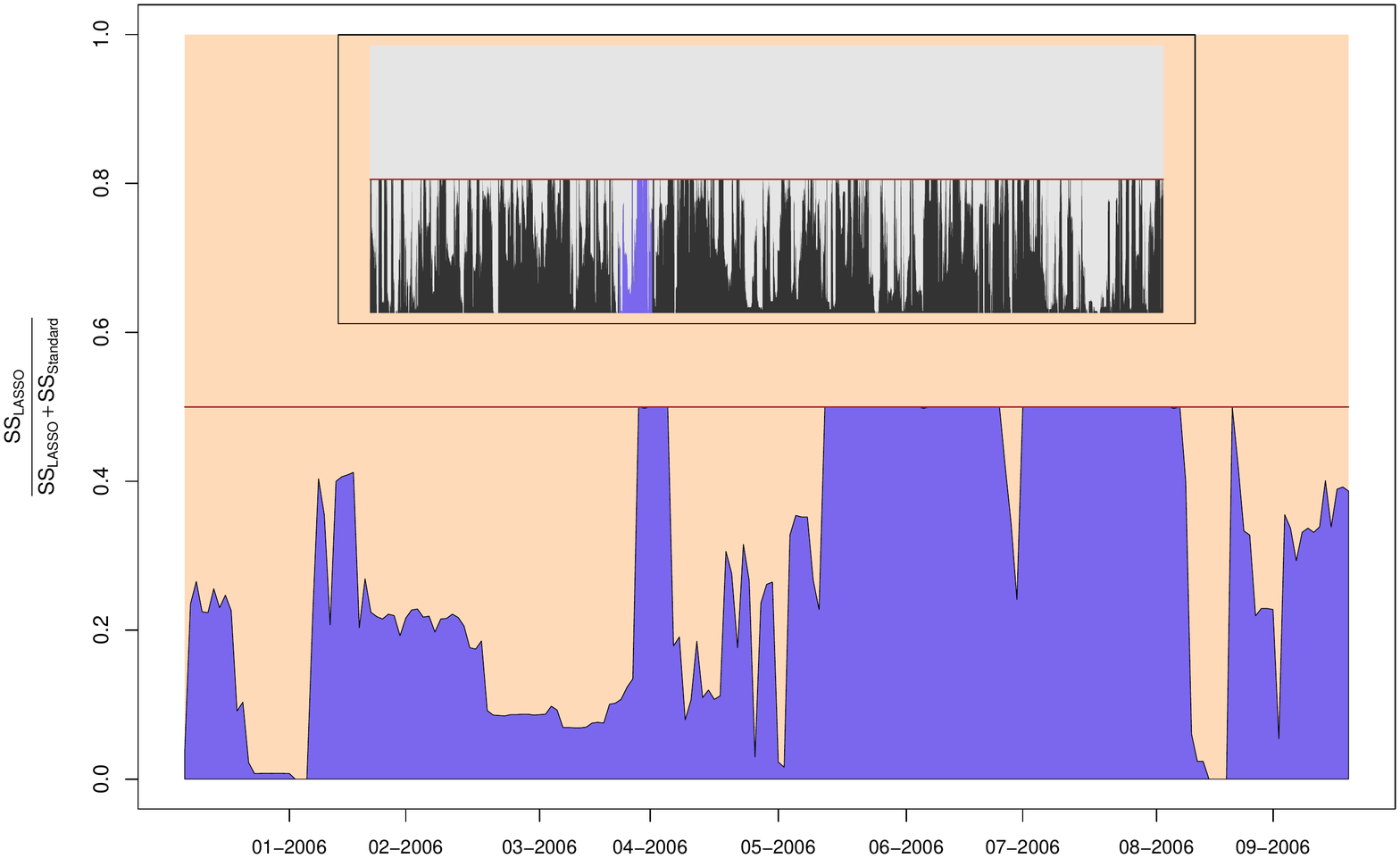}
  \caption{Comparison of short sales for the LASSO and Standard model.}
  \label{fig:sub1_ss}
\end{subfigure}%

\begin{subfigure}[b]{0.9\textwidth}
  \includegraphics[width=1\linewidth, height=8cm]{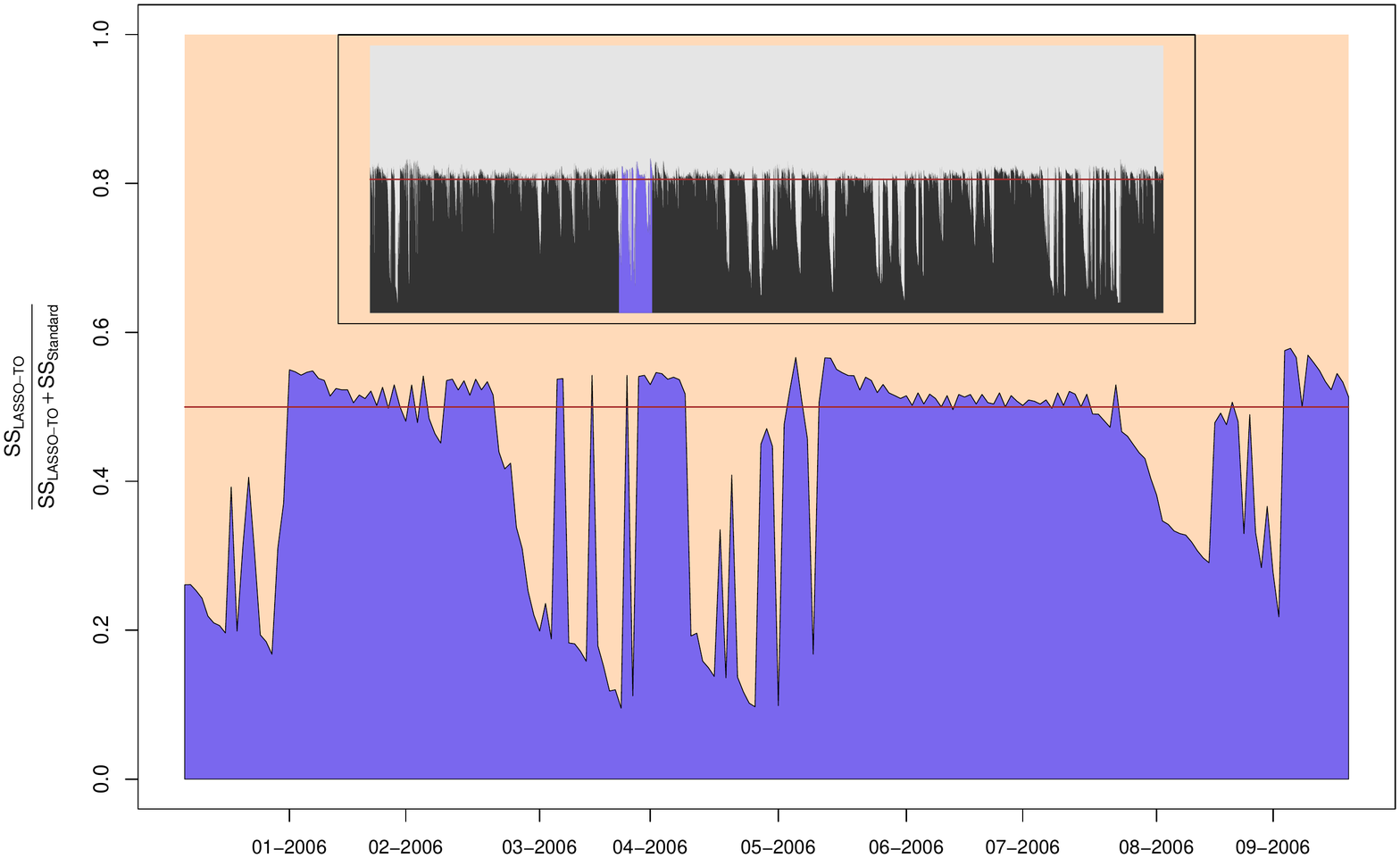}
  \caption{Comparison of short sales for the LASSO with the turnover constraint and Standard model.}
  \label{fig:sub2_ss}
\end{subfigure}
\caption{Development of short sales compared with the Standard model over the whole out-of-sample period as the black area and a snapshot of 10 months as the blue area. Data are normalized, so that 0 stands for 0 percent of the combined short sales of Standard and LASSO(+TO) and 1 for 100 percent. The red line describes the breaking point of 0.5, where both methods show the same amount of short sales.}
\label{fig:shortplot_nonlin}
\end{figure}

The fourth and last measure we investigate is average short sales, calculated as described in \eqref{eq_perf_measures_avg_short sales}. In accordance with the literature and as already seen in Figure \ref{fig:deltaplot_nonlin}, the LASSO tends to constrain short sales, leading to a small number of short sales overall compared with the Standard approach. In most cases, LASSO can generally halve the amount of short sales of the Standard approach, whereas the LASSO with the turnover constraint model induces a smaller reduction than LASSO alone. However, the LASSO with the turnover constraint model still decreases the short-sale budget overall compared with the Standard method. Figure \ref{fig:shortplot_nonlin} provides further details.

This illustration compares the amount of short sales of the LASSO-based model types with that of the Standard model type over time. The black and gray areas of the two subfigures represent the full out-of-sample period, whereas the blue areas are an example of a chosen subset to provide a better illustration. The figure specifically compares the proportion $\frac{SS_{LASSO(+TO)}}{SS_{LASSO(+TO)}+SS_{Standard}}$, where $SS$ stands for the amount of short sales on a specific day for a specific portfolio. This number is beneficial to analyze, as it explains which model exhibits higher amounts of short sales for a given timepoint. The number 0.5 stands for the cut-off point, for which one or the other model will have higher short sales, assuming that lower short sales is advantageous.

Subfigure \ref{fig:sub1_ss}, which compares the LASSO model with the Standard model, shows that the LASSO model never exhibits more short sales than its Standard version throughout the period. This is graphically illustrated by the black and blue areas never exceeding the red line. On some occasions (e.g., some days in August 2006) the short-sale budget of the LASSO model becomes 0 (i.e., there were no short sales).

Subfigure \ref{fig:sub2_ss}, which compares the LASSO with the turnover constraint model with the Standard model, does not share the same properties. Here, the red line is often exceeded by the black and blue areas, leading to a portfolio with higher short sales than under the Standard method. Moreover, short sales were never absent from the portfolio. In general, as already shown in Table \ref{tab_319}, the LASSO with the turnover constraint model still reduces the short-sale budget, as the gray area below the red line is greater than the black and blue areas above the red line.

	 \begin{table}[h!]
\begin{tabular}{  c  c  c  c  c  }

	 & ML & \citet{Ledoit.2003} & \citet{Ledoit.2017} & \citet{Fan.2013} \\ \hline
	\multicolumn{5}{c}{Standard deviation p.a.}  \\ \hline
	Standard & 0.1170 & 0.1137 & 0.1131 & 0.1136 \\ 
	LASSO & 0.1123 & \textbf{0.1116} & \textbf{0.1116} & \underline{\textbf{0.1112}} \\ 
	 & (0.0014) & (0.0703) & (0.1314) & (0.0051) \\ 
	LASSO $+$ TO & \textbf{0.1122} & 0.1119 & 0.1116 & 0.1113 \\ 
	 & (0.0000) & (0.0407) & (0.0643) & (0.0143) \\ \hline
	\multicolumn{5}{c}{Turnover}  \\  \hline
	Standard & 0.1348 & 0.1041 & 0.0922 & \underline{\textbf{0.0504}} \\ 
	LASSO & 0.2111 & 0.1829 & 0.1633 & 0.1247 \\ 
	LASSO $+$ TO & \textbf{0.0878} & \textbf{0.0858} & \textbf{0.0864} & 0.0730 \\ \hline
\multicolumn{5}{c}{Average assets}  \\  \hline
	Standard  & 100.00 & 100.00 & 100.00 & 100.00 \\ 
	LASSO & \underline{\textbf{56.61}} & \textbf{59.13} & \textbf{63.60} & \textbf{62.28} \\ 
	\textit{Percentage of full}  & \textit{56.61} & \textit{59.13} & \textit{63.60} & \textit{62.28} \\ 
	LASSO $+$ TO & 83.43 & 81.71 & 85.35 & 73.99 \\
	\textit{Percentage of full}  & \textit{83.43} & \textit{81.71} & \textit{85.35} & \textit{73.99} \\ \hline
\multicolumn{5}{c}{Average short sales}  \\  \hline
	Standard & 44.49 & 45.77 & 41.11 & 41.27 \\
	LASSO & \underline{\textbf{20.17}} & \textbf{22.62} & \textbf{20.97} & \textbf{21.12} \\ 
	\textit{Percentage of full}  & \textit{20.17} & \textit{22.62} & \textit{20.97} & \textit{21.12} \\ 
	LASSO $+$ TO & 37.90 & 37.42 & 36.20 & 30.06 \\ 
	\textit{Percentage of full}  & \textit{37.90} & \textit{37.42} & \textit{36.20} & \textit{30.06} \\ \hline
 \end{tabular}
	 \caption{Standard deviation p.a., average turnover per day, average assets as the mean of all non-zero weights, and average short sales as the mean of all weights greater 0 for the different models applied to the 100 S\&P dataset. The column headers represent the used variance estimation technique and rows report the results of the GMV calculated with \eqref{eq_general_frame_full1} (Standard), \eqref{eq_general_frame_full2} (LASSO), or \eqref{eq_general_frame_full3} (LASSO$+$TO). The results in bold represent the best model of that column for a specific measure and the underlined number represents the best overall model for that measure. The numbers in brackets represent the $p$-values of the test $H0: \sigma_{LASSO}=\sigma_{Standard}$ and $H0: \sigma_{LASSO+TO}=\sigma_{Standard}$, respectively. All the results are calculated based on the whole out-of-sample period consisting of 4778 daily observations.}
	 \label{tab_100}
	 \end{table}

To further strengthen our results, we examine 100 randomly selected stocks of the S\&P 500. Table \ref{tab_100} shows the results. The findings in Table \ref{tab_319} are supported by those summarized in Table \ref{tab_100}. \citet{Ledoit.2017} and \citet{Fan.2013} seem to be superior variance estimation techniques. Now, the standard deviation is almost always significantly better when using LASSO-based methods than Standard methods, providing even more evidence that our proposed methods can reduce the out-of-sample variance of the minimum-variance portfolio. Turnover for the POET Standard model is still slightly lower than that for the LASSO with the turnover constraint model simply because of our non-flexible choice of $k$ for all the models, as described above. This result can be easily adjusted by imposing a tighter turnover constraint parameter $k$. All the other major findings remain the same: the LASSO alone does not reduce turnover because of the need to estimate $\lambda$ for every period, but the LASSO with the turnover constraint model does. Furthermore, the LASSO models reduce the overall number of assets as well as the short-sale budget of the portfolios.

\section{Summary and conclusion}
In this study, we investigate different types of global minimum-variance portfolios in terms of their standard deviation and practically relevant features such as the number of included assets, a short-sale reduction, and a turnover constraint. We use realistic datasets with up to 319 stocks in one portfolio and find that highly efficient estimation techniques for minimum-variance portfolios can be combined with practitioners' requirements for such portfolios. Our proposed estimation setup is constructed to be easily implemented, as it is solvable with standard software for quadratic programming.

Adding common constraints found in the literature, we construct sparse and stable portfolios. Our detailed empirical analysis, covering almost 19 years of daily out-of-sample observations, show the distinct and novel features of using portfolio construction with efficient estimation techniques. Specifically, we make the following discoveries:
\begin{itemize}
\item LASSO-type models can retain the low-variance profile of highly efficient variance estimators or even lower it
\item A standard LASSO constraint increases turnover when $\lambda$ is allowed to change over time
\item The LASSO with the turnover constraint model can reduce turnover drastically, while maintaining sparsity, keeping variance low, and reducing the short-sale budget
\end{itemize}
We therefore conclude that it is beneficial, especially to practitioners, to use the LASSO with the turnover constraint approach and combine it with a modern technique to estimate large covariances. 

However, our results depend on the procedure used to obtain the lowest variance in the out-of-sample study, namely, the cross-validation for the different $\lambda$ tuning parameters. While this is a common practice in many fields unrelated to portfolio optimization, some researchers tend to use preset parameter values (i.e., they do not allow them to change over time). Future research should aim to provide more evidence on whether one or the other method yields better results in terms of relevant performance measures such as variance.

Further, data with a daily frequency are an a priori assumption that latently influences the outcome of our study. Even if the investor decides to rebalance daily, it is still questionable whether using only end-of-day daily data is sufficient to estimate the necessary moments of the multivariate return distribution. Advancements in intraday data analysis might therefore also be included in further research on this topic.

Finally, all our models inherit the assumption of homoscedasticity. This, however, conflicts with the inclusion of a time-varying $\lambda$ to some extent. Variance models that can capture the time dependency in return data, such as DCC-GARCH models, might be able to overcome the need to allow $\lambda$ to fluctuate over time. Recent advancements in combining non-linear variance estimation with time-dependent techniques might therefore also provide new insights into this matter.

\bibliographystyle{apalike} 
\bibliography{portfolio_V03} 

\clearpage

\end{document}